\documentstyle[bbm,amsfonts,epsfig,12pt,here]{article}

\newcommand {\eqref} [1] {(\ref {#1})}
\newcommand {\slsh} [1] {\not{\hbox{\kern-2pt${#1}$}}}

\newcommand{\drawsquare}[2]{\hbox{%
\rule{#2pt}{#1pt}\hskip-#2pt
\rule{#1pt}{#2pt}\hskip-#1pt
\rule[#1pt]{#1pt}{#2pt}}\rule[#1pt]{#2pt}{#2pt}\hskip-#2pt
\rule{#2pt}{#1pt}}

\newcommand{\Yfund}{\raisebox{-.5pt}{\drawsquare{6.5}{0.4}}}
\newcommand{\Yasymm}{\raisebox{-3.5pt}{\drawsquare{6.5}{0.4}}\hskip-6.9pt%
                     \raisebox{3pt}{\drawsquare{6.5}{0.4}}%
                    }
\newcommand{\Ysymm}{\Yfund\hskip-0.4pt%
                    \Yfund}

\def\drawbox#1#2{\hrule height#2pt
        \hbox{\vrule width#2pt height#1pt \kern#1pt
              \vrule width#2pt}
              \hrule height#2pt}

\def\Asym#1#2{\vcenter{\vbox{\drawbox{#1}{#2}
              \kern-#2pt       
              \drawbox{#1}{#2}}}}

\newcommand {\beq} {\begin{equation}}
\newcommand {\eeq} {\end{equation}}
 \newcommand {\ber}{\begin{eqnarray*}}
 \newcommand {\eer} {\end{eqnarray*}}
\newcommand {\bea}{\begin{eqnarray}}
 \newcommand {\eea} {\end{eqnarray}}

\newcommand{\Ntwo}{${\cal N}=2\ $}
\newcommand{\None}{${\cal N}=1\ $}

\def\Acknowledgements{\bigskip  \bigskip {\begin{center} \begin{large}
             \bf ACKNOWLEDGMENTS \end{large}\end{center}}}

\begin{document}

\begin{titlepage}
\begin{flushright}{CERN-TH/2003-087

TPI-MINN-03/10, UMN-TH-2135/03
}
\end{flushright}
\vskip 0.8cm

\centerline{\Large \bf  On \boldmath{$k$}-String Tensions and Domain Walls}
\vskip 0.1cm
\centerline{{\Large \bf in \boldmath{\None} Gluodynamics}} 

\vskip 1cm
\centerline{\large A. Armoni ${}^a$ and M. Shifman ${}^{a,b}$}
\vskip 0.1cm
\centerline{\small adi.armoni, michael.shifman@cern.ch}
\vskip 0.4cm
\centerline{${}^a$ Theory Division, CERN}
\centerline{CH-1211 Geneva 23, Switzerland}
\vskip 0.4cm
\centerline{${}^b$ William I. Fine Theoretical Physics Institute, University
of Minnesota,}
\centerline{Minneapolis, MN 55455, USA$^\star$}
\vskip 0.8cm

\begin{abstract}

We discuss the  $k$ dependence of the $k$-string tension $\sigma_k$ 
in SU($N$) supersymmetric gluodynamics. As well known, at large 
 $N$ the $k$-string consists, to leading 
order, of $k$ noninteracting fundamental strings, so that
$\sigma_k =k\sigma_1$. We argue, both from field-theory
and string-theory side, that subleading corrections to this formula run in
powers of $1/N^2$ rather than $1/N$, thus excluding the Casimir scaling. 
We suggest a heuristic model allowing one
to relate the $k$-string tension in four-dimensional gluodynamics
 with the tension of the
BPS domain walls ($k$-walls).
In this model  the domain walls are made of a net of strings
 connected to each other by baryon vertices.
The relation emerging in this way
leads to the sine formula $\sigma _ k \sim \Lambda^2 N \sin \pi k/N$.
We discuss possible corrections to the sine law, 
and present arguments that they are 
suppressed by $1/k$ factors. 
We explain
why the sine law does not hold in two dimensions. Finally, we discuss the
applicability
 of the sine formula for non-supersymmetric orientifold field theories.

\end{abstract}

\vspace{0.5cm}

\noindent
\rule{2.4in}{0.25mm}\\
$^\star$ Permanent address.
\end{titlepage}

\section{Introduction}
\label{secone}

\noindent

In confining theories, such as the Yang-Mills theory, non-supersymmetric or
supersymmetric ({\None} gluodynamics),  
heavy probe quarks are connected by color flux tubes. These
``QCD strings'' are the fundamental strings of the ``old'' string
theory of hadrons. A major improvement in our understating of
dynamics of  the QCD strings, due to the AdS/CFT correspondence
(Refs.~\cite{AdSCFT,Wilson}, for earlier ideas see Ref.~\cite{Polyakov:1997tj}), involves a
description in curved space of at least five dimensions.

\begin{figure}[H]
  \begin{center}
\mbox{\kern-0.5cm
\epsfig{file=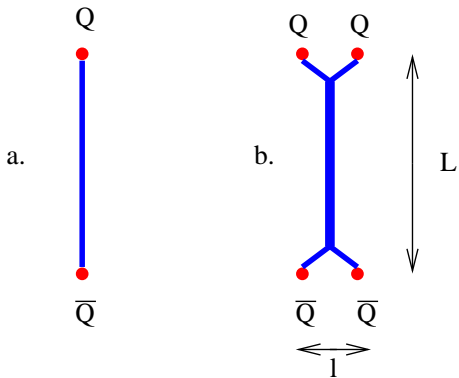,width=6.0true cm,angle=0}}
  \end{center}
\caption{A flux tube for $k$-strings. On the left (1a) is the
``fundamental'' tube. On the right (1b) is the 2-string tube.}
\label{st1}
\end{figure}

In this work we will consider confining gauge theories in four dimensions.
In what follows the gauge group is assumed to be SU($N$), and we will consider 
the limit of large $N$. Figure \ref{st1}a schematically shows
a flux tube connecting an infinitely heavy quark in the {\em fundamental} 
representation with its
antiquark. If $k$ probe quarks in
the  fundamental representation
are placed close to each other, a flux tube which connects them with
$k$ antiquarks is called a $k$-string, see Fig. \ref{st1}b displaying the $k=2$ example.

The tension of the $k$ string, $\sigma_k$, may be viewed as one of key parameters
of the confinement dynamics. It is under intense scrutiny since mid-1980's.
Most frequently discussed are two competing hypotheses:
(a) the Casimir scaling and (b) the Douglas-Shenker sine formula
(for extensive reviews and representative list
of references see e.g. \cite{one,two,matt}). The Casimir scaling hypothesis
reads that the $k$ dependence of $\sigma_k$ is
\beq
\sigma_k =\Lambda^2 \,k\, \left(1-\frac{k-1}{N-1}
\right)\,,\qquad k=1,2,...,N \,,
\label{csh}
\eeq
where $\Lambda$ is the dynamical scale parameter.
The sine  formula can be presented as follows: 
\beq
\sigma_k =N\, \Lambda^2\, \sin \left(\frac{\pi\,k}{N}  \right)\,,\qquad k=1,2,...,N \,.
\label{dsf}
\eeq
Both exhibit the symmetry
under the replacement $k\to N-k$, which corresponds to replacing quarks
by anti-quarks. Moreover,
both expressions imply  that  at $N=\infty$
\beq
\sigma_k = k\, \sigma_1\,,
\label{loosily}
\eeq
where $\sigma_1$ is the tension of the fundamental string (i.e. $k=1$).
The difference is in the subleading in $1/N$ corrections.
In the first case they run in powers of $1/N$ while in the second
in powers of $1/N^2$.

The purpose of this paper is two-fold. First, we will 
argue\,\footnote{Strictly speaking the
statement below refers to $d>2$. Two-dimensional gauge theories
must be discussed separately, see Sect. \ref{secfive}.
In two-dimensional
 QCD with {\em massive matter} the sine in Eq.~(\ref{dsf}) is replaced \cite{Armoni:1997ki}
by sine squared, namely,  $\sigma_k \sim m e  \sin ^2 \pi {k/ N}$. },
both from field theory and string theory side,  that corrections to Eq.~(\ref{loosily})
run in powers  of $1/N^2$. This then leaves no space to the Casimir
scaling in four dimensions.
Second, we suggest a heuristic picture, based on $k$-walls in \None gluodynamics,
from which Eq.~(\ref{dsf}) naturally follows. 

Domain walls are BPS objects in \None gluodynamics. 
They interpolate between distinct vacua of the theory, say
vacuum $n$ and vacuum $n+k$ (altogether there are $N$ vacua).
The domain walls interpolating between neighboring vacua are called ``elementary,''
while at $k>1$ we deal with the $k$-walls which can be viewed as bound states
of $k$ elementary walls.

The tension of $k$-walls is known  {\it exactly}  \cite{Dvali:1996xe},
\begin{eqnarray}
T_k &\equiv&  T_{n,n+k} = \frac{N}{8\pi^2}\, 
\left|\langle \lambda \lambda\rangle_{n+k}-  \langle\lambda \lambda\rangle_{n}\right|
\nonumber\\[2mm]
&=&
N^2 \, \Lambda ^3 \, \sin \left(\frac{\pi\,k}{N}  \right)\,,\qquad k=1,2,...,N \,, 
\label{wall-tension}
\end{eqnarray}  
(note that $T_{n,n+k}$ is $n$-independent).
The similarity between Eqs.~(\ref{dsf}) and (\ref{wall-tension})
is apparent. This becomes more than a mere similarity if one accepts
our qualitative picture:
we suggest that  $k$-walls  can be thought of as being ``composed'' of 
fluctuating $k$-strings
 connected by baryon vertices (junctions of $N$ strings). This picture
originated from the observation that the wall tension scales as
$N$, at large $N$. Moreover, it is supported by the recent finding
\cite{AV} that a level $N$ Chern-Simons theory lives on the wall. 

Indeed, if our picture is correct, one can easily show that 
\beq
T_{k} \sim  \sigma _k \, ( \Lambda \,  N ) + \mbox{subleading corrections}\, , 
\label{wall-string}
\eeq
which leads directly to Eq.~(\ref{dsf}) for the $k$-string tension.
We can also reverse the logic: if Eq.~(\ref{dsf}) is indeed a good
approximation to the $k$-string tension of \None gluodynamics --- a
$k$-wall can be well approximated by a net of $k$-strings.

The organization of this paper is as follows: in Sect. \ref{sectwo} we explain
why SU($N$) gauge dynamics leads to $1/N^2$ corrections to the leading
$k$-string tension. In Sect. \ref{secthree} we present our model and  provide
supporting
arguments both from field theory and string theory. 
Section \ref{dopoln} is devoted to analysis 
of the $k$ dependence of binding forces
versus $N$ dependence.
A remark concerning non-supersymmetric
Yang-Mills theories is presented in Sect. \ref{secfour}. 
In  Sect. \ref{secfive} we
explain why the two dimensional case and the strong coupling lattice
expansion are special. Section \ref{secsix}  is devoted
to concluding remarks.

\section{Corrections: \boldmath{$1/N$} or \boldmath{$1/N^2$}? }
\label{sectwo}

\noindent

At $N=\infty$ both $k$-walls and $k$-strings present {\em ensembles}
of $k$ noninteracting constituents --- elementary walls and 
fundamental strings, respectively.
The binding emerges as a subleading in $1/N$ effect.
The Casimir scaling predicts it at level $1/N$ while the sine formula
at level $1/N^2$. In both cases the binding is weak at large $N$.

The Casimir scaling was abstracted from various models
in which confinement is due to a one-gluon 
exchange\,\footnote{The quadratic Casimir coefficient
$C_R$ is defined as $T^aT^a = C_R \times 1_R$ where $T^a$ stands for the SU($N$)
generators in the representation $R$ while $1_R$ is the unit matrix in the 
same representation. The color structure of the one-gluon exchange is proportional
to $C_R$.} (perhaps, appropriately modified in the infrared domain).
In Sect. \ref{secfive} we discuss an example --- two-dimensional
Yang-Mills theory.
It enjoyed considerable numerical support from lattice
simulations \cite{casc}. It is also obtained in the lattice strong
coupling ($1/g^2$) expansion.    

The sine law was motivated by various investigations
of the Seiberg-Witten model and, almost simultaneously,
 from the string theory side.
Equation (\ref{dsf}) was first derived by Douglas and Shenker \cite{Douglas:1995nw}
 as the QCD string tension in the softly broken \Ntwo
Yang-Mills theory\,\footnote{The original
Douglas-Shenker formula  is $\sigma _k = N\, m\, \Lambda\, 
\sin \left({\pi\,k}/{N}  \right)$.
Corrections of the order  $O(m^2)$  to this expression
were studied in Ref.~\cite{ken}. Note that the adjoint mass parameter
$m$ in Ref.~\cite{Douglas:1995nw} differs from that in Ref.~\cite{ken}
by a factor of $N$. The large $N$ scaling in softly broken \Ntwo
is not straightforward since the lightest ``$W$ boson'' masses are 
not $N$ independent; in fact, they are of the
order $O( \Lambda/N^2 )$. At $N\to\infty$ the description
of the low-energy physics based on the 
U(1)$^{N-1}$ limit inherent to the Seiberg-Witten solution,
becomes invalid. The condition of applicability of Eq.~(23) in 
Ref.~\cite{ken} is $\, m\, N \ll\Lambda $.
}.
Then it was  obtained in the context of MQCD \cite{Hanany:1997hr} and, more
recently,
in the AdS/CFT framework \cite{Herzog:2001fq}. In the latter
case, the sine formula (\ref{dsf})  was
found to be exact for the Maldacena-Nu\~{n}ez background
\cite{Maldacena:2000yy} and valid to a few percent accuracy 
in the Klebanov-Strassler
 background \cite{Klebanov:2000hb}. Recent lattice simulations 
suggest \cite{DelDebbio:2001kz,Lucini:2001nv} that 
the sine formula  fits the
$k$-string tension in pure Yang-Mills theory in four dimensions
better than the Casimir formula.

Here we would like to address an aspect crucial for discriminating between
the Casimir and  sine laws, namely the $N$ dependence of the binding
energy of $k$ fundamental strings. For simplicity we will use
the example with $k=2$, although the argument is general and applicable 
for any $k$.

\begin{figure}[H]
  \begin{center}
\mbox{\kern-0.5cm
\epsfig{file=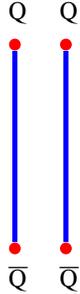,width=1.0true cm,angle=0}}
  \end{center}
\caption{Two free ``fundamental'' flux tubes.}
\label{st2}
\end{figure}

Start from two well-separated fundamental quark-antiquark pairs, as 
it is shown in Fig. \ref{st2}.
When the separation $l$ 
between two $Q$'s (or, which is the same, between two $\bar Q$'s)
is large enough it is obvious
that the tension of this configuration is just twice the fundamental
tension $\sigma_1$. 
As we adiabatically move the pair on the left towards the pair on
the right, the attraction between the flux tubes switches on, and at
$l < L/N$ ($L$ is the length of the string, see Fig.  \ref{st1}b) the 
2-string configuration of Fig.  \ref{st1}b becomes
energetically favorable, see Appendix for a detailed derivation. When 
the separation $l$ becomes
$<\Lambda^{-1}$  all remnants of 1-strings disappear. A similar
process, described via supergravity, is given in \cite{Gross:1998gk}.
The quarks
$Q^i\,,\,\, Q^j$ (as well as $\bar Q_k\,,\,\, \bar Q_\ell$)
are in the mixed color state, symmetric plus antisymmetric,
but this is unimportant because the $k$-string tension is supposed 
to depend only on the
$N$-ality. It is perfectly consistent to consider the {\em reducible} 
two-index representation. 

Now, we can proceed to analysis of
the graphs responsible for the attraction of two 1-strings.
First of all, it will be convenient to formally define the
2-string tension (the definition for $k$-string is similar).
Consider two rectangular Wilson contours, $C_1$
and  $C_2$, both lying in the $\{x\,t\}$ plane and separated by
distance $l$ in the $y$ direction ($z=0$). The size of both contours is
$L\times T\to\infty$, and they are parallel to each other.
Each contour gives rise to the Wilson loop operator
\beq
{\cal W}[C_i] = {\rm tr}\, \exp \left(i \int_{C_i} A_\mu \, dx_\mu \right)\,,\quad i=1,2\,,
\label{wilop}
\eeq
describing the time evolution of 1-string
of length $L$. Two 1-strings are parallel, oriented in
the $x$ direction  and separated by an interval $l$ in the $y$ direction;
$z$ is set to zero.
Then we define
\beq
\langle {\cal W}[C_1]\,,\,\, {\cal W}[C_2]\rangle = \exp\left(-\Sigma_2(l) \, T\, L
\right)\,.
\eeq 
At $l\gg\Lambda^{-1}$
$$\Sigma_2(l) \to 2\, \sigma_1\,,$$
while at  $l\ll\Lambda^{-1}$
$$\Sigma_2(l) \to  \sigma_2\,.$$

\begin{figure}[H]
  \begin{center}
\mbox{\kern-0.5cm
\epsfig{file=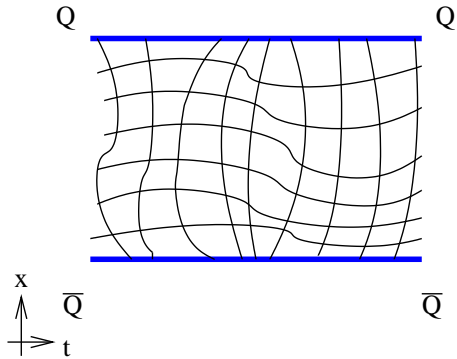,width=6.0true cm,angle=0}}
  \end{center}
\caption{A ``fishnet'' --- the flux tube worldsheet. Thick lines denote
heavy probe (anti)quarks in the fundamental representation of SU($N$).}
\label{fish1}
\end{figure}

\begin{figure}[H]
  \begin{center}
\mbox{\kern-0.5cm
\epsfig{file=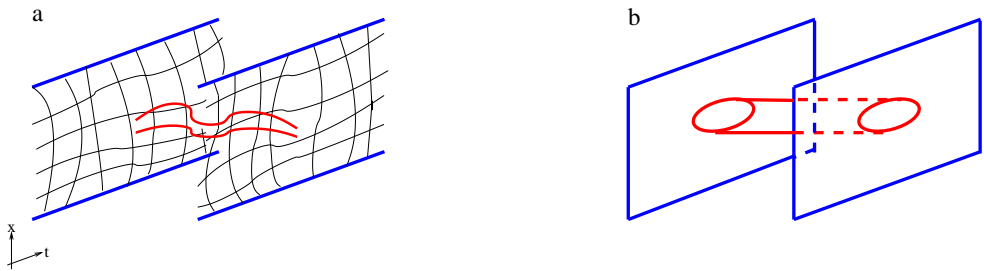,width=13.0true cm,angle=0}}
  \end{center}
\caption{The interaction of two ``fundamental'' flux tubes: (a) 
Field-theory picture --- two-gluon exchange; (b) String-theory picture --- 
exchange of a closed string between two worldsheets.}
\label{fish2}
\end{figure}

In field theory language the fundamental QCD string 
evolving in time can be viewed as a ``fishnet''
(Fig. \ref{fish1}). The attraction of the flux tubes is due to gluon exchanges connecting
two planar ``fishnets'' (Fig. \ref{fish2}a). Note that the minimal
number of gluons that are exchanged between the worldsheets is two,
since one cannot transfer color between the flux tubes. It is rather obvious that
the gluon exchanges
are suppressed by $1/N^2$ since this is a non-planar contribution.
Note that even gluons coupled to probe quarks do not produce $1/N$.
Correspondingly, the sum of two Casimir coefficients,
$C_{\Yasymm} + C_{\Ysymm}$,  has no $1/N$.
The factor $1/N$ occurs at the stage of division of the reducible representation
into two irreducible, $\Yfund\otimes \Yfund = \Yasymm \oplus \Ysymm$.
This separation is irrelevant for our purposes.
The 1-string attraction (per unit length)
is determined by exchanges that are localized in space
on the scale $\Lambda^{-1}$.
How the middle parts of the long  1-strings interact is independent
on  details of endpoints which are separated
from the middle part by distance $L\gg \Lambda^{-1}$ 
 --- whether the quarks are symmetrized or antisymmetrized
with respect to color, or none, does not matter.
It must depend only on $N$-ality. In the general case of the $k$-string
the quadratic Casimir is
\beq
C_R = kN + \sum _i r_i^2 - \sum _i c_i^2 - {k^2 \over N},
\eeq
where $r_i$ and $c_i$ are the lengths of the rows and the columns, respectively,
of the Young tableau of the representation $R$. As we 
have argued above for the $2$-string, in  the general
case the string tension should depend only on the 
$N$-ality and not on the specific representation --- hence, it cannot depend
on $r_i$ or $c_i$ which  lead to  potential $1/N$ dependences. 

When
the strings are separated by a large distance, $l\gg \Lambda^{-1}$, 
it makes no sense to speak of the gluon exchanges
between the fundamental strings.
The force between
well-separated 1-strings 
 is controlled by the lightest glueball exchange
(``scalar dilaton''), which is certainly $1/N^2$ effect.  

In the AdS/CFT framework, the Wilson loop is described by a minimal
surface \cite{Wilson} (the string worldsheet) that extends ``inside'' the AdS space,
\beq
\langle{\cal W}\rangle=\exp\left( -S_{\rm NG}\right) ,
\eeq
where $S_{\rm NG}$ is the Nambu-Goto action. The $k$-string is described by $k$
coincident elementary worldsheets \cite{Gross:1998gk}. Clearly, the
string tension will acquire a factor $k$ at the level of free strings,
 $\sigma _k = k \sigma _1$. The interaction
between string worldsheets in the bulk AdS is via an exchange of closed
strings (see Fig. \ref{fish2}b) and this process is obviously proportional to
$g_{\rm st}^2$ which translates into $1/N^2$. In fact, string theory
predicts not only the right power ($1/N^2$), but also the right sign
(minus): the
flux tubes will attract due to an exchange of NS-NS fields.

\section{The model}
\label{secthree}

\noindent

In this section we outline  our model or, better to say, a heuristic picture
which has been mentioned in the introduction. We use
both field theory and string theory languages.

\begin{figure}[H]
  \begin{center}
\mbox{\kern-0.5cm
\epsfig{file=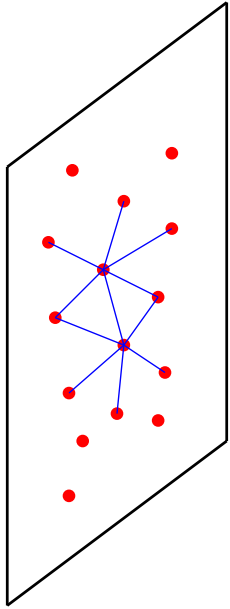,width=3.0true cm,angle=0}}
  \end{center}
\caption{The domain wall 
in supersymmetric gluodynamics
as a net of strings joined via  baryon vertices.
The example above refers to SU(6). One should imagine an irregular lattice of nodes connected
by lines in a chaotic (but planar) way, so that each node either receives or emits $N$ lines. The net fluctuates in the quantum-mechanical sense,
and in no way represents a regular structure similar to fullerenes.}
\label{wall}
\end{figure}

\subsection{Field theory picture}
\label{secthreeone}

\noindent

The elementary  wall tension 
in \None gluodynamics scales  at large $N$
as $N\Lambda^3$. 
One may ask how the factor $N$ appears in the theory
where there are no fields in the fundamental representation of SU($N$),
only in the adjoint. 
Ordinary solitons scale as $N^2$. This led Witten \cite{Witten:1997ep} to 
suggest that the BPS domain walls at hand are in fact  QCD D-branes. 
A   D-brane signature of the \None domain walls
is that  QCD strings can end on these domain walls
\cite{Witten:1997ep} (see also \cite{Kogan:1997dt,Gabadadze:1999pp,GS})
and parallel walls can exchange glueballs (closed
strings) between them \cite{Armoni:2003jk}.

This is no explanation on the microscopic field-theoretic level, however.
In a bid to explain the linear $N$ dependence of the 
wall tension in a ``natural way'' we will assume that
the wall presents a (fluctuating) network of interconnected
$N$-string junctions, as in Fig.~\ref{wall}. It is known since  long  that
$N$ fundamental strings can join each other in a baryon vertex.
Let us assume that these nodes form a 
flat two-dimensional structure.
Every node either emits or absorbs $N$ fundamental strings.
The strings intertwine the nodes randomly (but in a planar way)
creating an elementary domain wall. If the density of nodes is of order one per
area $\Lambda^{-2}$ of the wall surface, then the tension will naturally scale as
$N$. 

Since the wall is made of $N$-string junctions
we can relate the tension of the wall to the tension of the string.
Indeed, on the one hand, the energy of the square with area $\Lambda^{-2}$
on the elementary wall is $T_1 \Lambda^{-2}$, by definition.
On the other hand, it is of the order of $\sigma_1\, N\,\Lambda^{-1} $.
Thus, $\sigma_1\sim T_1 \,( N\,\Lambda )^{-1}$. The axial charge of
the wall is carried by the baryonic junctions (we discuss this issue
in more details in the next subsection). 

Now, let us pass to $k$-strings and $k$-walls.
It is established that $k$ elementary walls
form a bound state, the $k$-wall \cite{adam,rsv}.
It is also well established that the multiplicity of the $k$-wall is \cite{AV,rsv}
\beq
\nu_k = \frac{N!}{k! (N-k)!}\,.
\label{multiplicity}
\eeq
This is also the multiplicity of the {\em antisymmetric} $k$-index representation.
On the other hand, as various arguments indicate,
for $N$-ality $k$
the lowest energy configuration for two distant groups of 
$k$ fundamental quarks (antiquarks) corresponds to quarks in the
{\em antisymmetric} representation of SU($N$).
This is suggestive that the $k$-wall is built of the 
$k$-strings in the same way as the elementary wall of fundamental strings.
The $k$ wall can be viewed as  a flat two-dimensional 
fluctuating web of nodes
each of which emanates or absorbs $N$ $k$-strings.
Here we assume the existence of the $k$-string junctions, which can be viewed
as bound states of 1-string junctions.

If so,  we arrive
at the relation (\ref{wall-string}). 
Unlike the $k$-wall tension which is given by the sine formula {\em exactly},
see Eq.~(\ref{wall-tension}), we do not expect the sine formula to be exact for
the $k$-strings:
 the wall is a BPS object, whereas the QCD string is not.
 
A general function that satisfies the requirements: (i) 
symmetry under  $k\rightarrow N-k$;
(ii) being even in powers of $N$;  and (iii)
 $\sigma _k $ tending to $k\sigma_1$  at large $N$, can be written as follows:
\beq
N \Lambda ^2 \sum _{{\rm odd\,\, } l} C_l (N)\left (  \sin  {\pi k\over N} \right
)^l
\, ,
\eeq
with $C_1 =1$. Since the QCD string is non-BPS, {\em apriori} all $C_l$ are
non-vanishing.

What can be said of corrections in the right-hand side of Eq.~(\ref{wall-string})?
In the web of $k$ strings forming a $k$-wall, 
 binding interactions in the direction
transverse to the wall plane, which are $k$-dependent,
are contaminated by interactions along the wall plane, which we 
expect to be $k$ independent. That is to say that we expect 
$$
{\sigma _k \over k}
- \Lambda ^2 \left( {N\over k}\right)  \sin \frac{\pi \,k}{ N} = O(1/k)\,.
$$
If so, the corrections  in the right-hand side of Eq.~(\ref{wall-string})
are of order $1/k$. At large $N$ and $k/N$ fixed, these corrections are then subleading.

We would like to mention a possible difficulty in our model,
a question to be addressed in the future\,\footnote{A.A. thanks
J. Barb\'on for addressing both the problem and its possible solution.}.
The 1-wall thickness
is $\sim 1/ (\Lambda N)$ \cite{GS},
whereas the flux-tube thickness is much larger,
 $\sim 1/\Lambda$. That is, the QCD 1-string hardly fits the 1-wall
in the perpendicular direction. The thickness of  $k$-walls and  $k$-strings
has not yet been  established. It might well be
that the thickness of both of them at $k/N$ fixed and $N\to \infty$
is the same. This is another
reason why our picture may be successful at large $k$ but definitely
fails at small $k$.

The arguments above certainly do not have the status
of a direct field-theoretical proof of
 Eq.~(\ref{dsf}) in  \None gluodynamics.
They are nevertheless supplementary and independent of other arguments,
such as MQCD-based derivation (it does
not apply to \None gluodynamics, but, rather, to unknown theories in the 
same universality class),
or the supergravity argument (which holds for theories  having a different
ultraviolet content than \None gluodynamics). 
The picture we have conjectured  is entirely  within field theory
{\em per se}.

\subsection{String theory picture}
\label{secthreetwo}

\noindent

From the string theory point of view our model is rather natural.
For example, in the Polchinski-Strassler supergravity dual \cite{Polchinski:2000uf},
domain walls are represented by wrapped D5 branes that look
like D2 branes from the 4d point of view. The baryons are wrapped D3 
branes that look effectively like D0 branes. It was even conjectured by
Polchinski and Strassler that ``an assembly of baryons can be arranged
into a spherical domain wall!''.

In the type IIA realization
of \None gluodynamics by Acharya and Vafa \cite{AV}, domain walls are
represented by wrapped D4 branes and baryons as wrapped D2 branes.

In both realizations  \cite{Polchinski:2000uf,AV} the wall looks like a D2 brane and the
baryon-vertex as a D0 brane. We suggest that the D2 brane is made
out of D0 branes and fundamental strings that connect them.

Indeed, according the Acharya-Vafa \cite{AV} there is a level $N$ 
Chern-Simons (CS) term on the wall. The source of this term is the interaction  
term of the D4 with a RR bulk field. Since the D4 wraps an $S^2$ with
a total RR flux of $N$ units, we get a level $N$ CS term in the 2+1
theory.

The CS term can be viewed as a baryon vertex, if the field
strength on the domain wall is replaced by a delta function. This is
exactly what we obtain by wrapping a D2 over $S^2$. So, the the
wrapped D4 can be thought of as made out of D2 branes: both serve as 
a source for $N$ fundamental strings.

To conclude, we suggest that the wall (a D-brane) is made out of a net
of lower dimensional D-branes and the strings that connect them. The
tension of the wall is due to the tension of the strings (but also due
to the tension of the D0 branes). 

The axial charge of the wall, that
is the RR charge of the D2 brane, is due to the RR charge of the
D0 branes (the baryon vertices).
This should explain why the order parameter
$\langle\lambda\lambda\rangle$ changes its phase
when one pierces the wall in passing from left to right or
{\em vice versa}.

 It is difficult to understand  how the
wall, which is a BPS object, is built from non-BPS constituents 
(recall that even the wrapped
D2 is not BPS). It means that our construction is approximate. At
large $k$, however, the corrections are expected to be small.

The picture that we advocate in this section is obviously similar to Myers 
effect \cite{Myers:1999ps}.

\section{\boldmath{$k$} dependence versus \boldmath{$N$} dependence}
\label{dopoln}

\noindent

In Sect.~\ref{sectwo} we discussed in detail the structure
of the $1/N$ corrections responsible for the binding of $k$ 1-strings into
a $k$-string. The heuristic picture of ``$k$-strings as building blocks for $k$-walls,''
discussed in detail in Sect.~\ref{secthree}, leads, at large $k$, 
to the sine law, Eq.~(\ref{dsf}),
for the $k$-wall tension, which implies not only a very special
$N$ dependence, but a special $k$ dependence too. 
This $k$ dependence may seem counterintuitive, at first sight.
Our task is to invert the argument and see whether we can learn anything new
from this analysis.

Let us start from the $k$-wall tension (\ref{wall-tension}), which is the exact result of
\None gluodynamics. (In the remainder of this section, for simplicity,
we put $\Lambda =1$.) Expanding the sine we get
\beq
T_k = N\pi \, k - \frac{\pi^3}{6}\,\frac{k^3}{N} + \frac{\pi^5}{120}\,
\frac{k^5}{N^3}-....
\label{sinexpa}  
\eeq
The first term represents $k$ noninteracting
1-walls, while the second and higher terms are due to a binding
force. Let us examine the second term, proportional to $k^3/N$.
The $N$ dependence tells us that it is due to a binary interaction,
since the domain walls are QCD D-branes, see Ref.~\cite{Armoni:2003jk}.
Naively one would then require
a coefficient  proportional to $k^2$ rather than $k^3$, which is just 
a combinatorial factor.
The exact formula tells us that this is impossible. What went wrong?

When we have an {\em ensemble}
of $k$ constituents with  binary interactions (with a coupling $g^2$)
bound in a ``compound nucleus,'' the binding energy scales
as $g^2\,k^2$ only provided that the size of the system and the
coupling $g$ are 
$k$-independent. This is what happens, for instance, in Witten's picture of 
baryons \cite{ewb}. However, in the case of the domain walls, we have an
ordered system along the line, with a non-universal 
coupling\,\footnote{We thank A. Ritz 
who pointed out to us the importance of the wall
ordering and non-universality of the interaction force.}.
 The coupling
of a wall to its $\ell$-th neighbor is $\ell/N$. The sum over
all possible pairs yields the $k^3$ behavior. A similar $k^3$ behavior
 is expected in a one-dimensional system with size $\sim k$ and a linear
potential between pairs. Other examples of one-dimensional ordered
$k$-body systems with the ground state energy scaling as $k^3$
are known in the literature, see e.g.
\cite{thou}.

As we pass to $\sigma_k$, the situation may  be rather similar.
In this case the sine law implies
\beq
\sigma_k = \pi \, k - \frac{\pi^3}{6}\,\frac{k^3}{N^2} + \frac{\pi^5}{120}\,
\frac{k^5}{N^4}-....
\label{sinexpas}
\eeq
Again, the $N$ dependence of the term
$k^3/N^2$ has the structure typical of binary interactions.
A ``natural''  $k$ dependence in this case, as was mentioned, is $k^2$.
This ``natural'' $k^2$ factor follows from a
picture of a ``compound state'' with a typical size being $k$-independent.
Basing on the lesson of $k$-walls we are
inclined to think that a more complicated $k^3$ scaling takes place due
to a $k$-dependence of a typical size of
 the cross-section of the $k$-string.

\section{What can be said about non-supersymmet\-ric gauge theories?}
\label{secfour}

\noindent

Above we presented a model yielding
 a relation between the $k$-wall and $k$-string tensions
in \None gluodynamics. It is natural to ask now
about  non-supersymmetric gauge theories, in particular,
the orientifold theory discussed in Ref.~\cite{asv}. 
In this theory, the gluino adjoint field is replaced by
a Dirac two-index field in the antisymmetric representation, which, obviously,
makes it non-supersymmetric.
The orientifold theory was shown \cite{asv} to be planar
equivalent to \None gluodynamics
in the mutually overlapping sectors. At large $N$ 
both have $N$ discrete vacua and  domain walls of a very similar structure.
This may suggest that the $k$-string tension in the
non-supersymmetric orientifold theory is described
by the same sine formula as in \None gluodynamics.

At first sight this suggestion looks heretical.
Indeed, in  \None gluodyna\-mics any $k$-index source cannot be screened,
and, therefore, develops a string.
On the other hand, in the orientifold theory,
any source with even $k$ can be completely screened by dynamical 
quarks. Any source with odd $k$ can be almost screened ---
only one fundamental index remains unscreened.
Common wisdom prompts one 
that there is no place for such quantity as $\sigma_k$ in this
theory.

Let us remember, however,  that our argument in favor of the
sine formula refers to large $k$, namely, $k\sim N$ at $N\to \infty$ (we can obtain no result
otherwise). At large $k$, in order to screen the
$k$-index source, one has to (pair)produce
 a large number of dynamical quarks, scaling as
$N$,  from the
vacuum.
This process is suppressed, presumably exponentially, as
$e^{-Ck}$ where $C$ is a positive number
(we are  certain that it is at least power suppressed).
Thus, at large $k$ there is confinement in the 
orientifold theory,  $\sigma_k\neq 0$, and, according to our
hypothesis, obeys the sine law.

\section{Why two-dimensional Yang-Mills theory \\
and strong coupling lattice expansion \\
are special?}
\label{secfive}

\noindent

In two-dimensional Yang-Mills theory the ``$k$-string tension''
is known exactly, and it
obeys the Casimir scaling,
 $\sigma = g^2 C_R$. 
A question which immediately comes to one's mind is ``what makes two-dimensional 
formula special?''

The answer is as follows. Genuine strings, with local interaction
(such as those one deals with in four dimensions) never develop in two dimensions.
The two-dimensional  Yang-Mills theory (in the axial gauge) is free, Gaussian.
Non-linearity is absent, and so are transverse dimensions. Confinement is 
due to the one-gluon exchange (instantaneous in time) which provides a linear potential
with the coefficient proportional to the Casimir operator $C_R$.
As a reflection of the absence of the {\em bona fide} strings in two dimensions,
please, observe that the ``$k$-string tension''
 depends in this case
on the specific representation and {\em not} merely  on the $N$-ality.

The same situation takes place in a variety of {\em ad hoc}
four-dimensional models where (a modified) one-gluon exchange
is postulated to be responsible for confinement.
In these models the gluon propagator at small momenta
is assumed to scale as $1/p^4$ rather than $1/p^2$.
This certainly provides a linear confinement with the $C_R$
coefficient. This approach is totally inconsistent
with our arguments.

The situation in (four-dimensional)
lattice strong coupling expansion
\cite{strcl} is similar to that in 
two-dimensional Yang-Mills theory.
As well-known, see e.g. \cite{one}, the strong coupling Wilson expansion
implies a linear confinement with the coefficient in front
of the linear potential determined by $C_R$. The ``1-strings''
 do not interact ``in the bulk.'' Therefore, we do  not have a {\em bona fide}
1-string binding of the type we expect in actual 
four-dimensional Yang-Mills theory, where the scale
parameter is dynamically generated, 
and the attraction of 1-strings (per unit length)
is localized along the string.

\section{Discussion and conclusions}
\label{secsix}

\noindent

The main thrust of this paper is on the model presenting the BPS $k$-wall in 
\None gluodynamics as a planar  network of baryonic vertices (nodes)
connected by $N$ $k$-strings. The network is irregular and fluctuating.
The model is admittedly heuristic and qualitative.
Taking it at  its face value, however, one can 
get a relation between the $k$-wall tension which is exactly
known, and the $k$-string tension.
 
Our relation implies the Douglas-Shenker sine formula.
Irrespective of particular models is the result
we have proven {\em en route}: the $1/N$ corrections to the free string limit,
$\sigma_k =k\sigma_1$, run in powers of $1/N^2$ rather than $1/N$.
This rules out the Casimir scaling.
In general,  models with linear confinement 
basically fall into two categories.
The coefficient in front of the linear term in the potential between 
two probe objects in the given model may either depend on a particular color representation
of the probe object, or only on the $N$-ality.
In the first class of models corrections to
$\sigma_k = k\,\sigma_1$ run in powers of $1/N$.
In the second class, which  includes four-dimensional Yang-Mills
theory, corrections run in powers of $1/N^2$. Note that
in the first class of models,
it is wrong to label the string tension only by the $N$-ality $k$. One should  indicate the
corresponding Young tableau for the particular color representation 
of the probe object.

The central question in the issue of the
$k$-string tension is the nature and size of corrections to the sine formula.
We presented an argument showing that at $k/N$ fixed and $N\to \infty$
these corrections are suppressed by $1/k$ factors, i.e.  suppressed
parametrically.
On the other hand, 
one of the backgrounds considered in Ref.~\cite{Herzog:2001fq}
(the Klebanov-Strassler background \cite{Klebanov:2000hb}) leads to corrections 
to the sine formula which are
strongly suppressed numerically, but not parametrically.
Is it due to the fact that the ultraviolet behavior of the AdS/CFT
models is different from that in \None gluodynamics?
The question remains open. Moreover, recently it was argued~\cite{Herzog:2002ss} 
that in (1+2)-dimensional Yang-Mills theory
corrections to the sine formula are so large that they, in fact, convert it
into a (well) approximated Casimir formula 
for the background of Cveti\v{c} {\em
et al.}  \cite{Cvetic:2001ma}, an analog of the Klebanov-Strassler
background. It is curious that
 for the (1+2)-dimensional analog of the Maldacena-Nu\~{n}ez
supergravity background 
(see Ref.~\cite{Maldacena:2001pb}) a sine formula is 
suggested in Ref.~\cite{Herzog:2002ss}. From our standpoint, the
only difference between
the (1+2)-dimensional and (1+3)-dimensional gauge theories
is due to the fact that in 1+3 dimensions the mass scale is dynamically
generated (via dimensional transmutation), while in 1+2 dimensions
the mass scale is given by the square of the
gauge coupling constant. In both cases non-linearities are well-developed,
and the spacial string structure is ``local.'' In other words,
the Casimir scaling behavior looks suspicious,
 and, in fact, we expect that in 1+2 dimensions 
large-$N$ corrections to $\sigma_k =k\sigma_1$
run in powers of $1/N^2$, much in the same way as in
 1+3 dimensions. 
 
It is important to note that our consideration is fully invertible.
If in the future someone succeeds in deriving the
Douglas-Shenker sine formula from other principles,
this will simultaneously substantiate our picture
of the BPS domain $k$-walls made of  fluctuating networks of $k$-strings.

\Acknowledgements

We thank L. Alvarez-Gaum\'e, J. Barb\'on, P. Di Vecchia, 
and G.~Veneziano for numerous discussions.
Very useful communications with C. Herzog, I.~Kogan, K. Konishi,
P. Rossi, A. Turbiner, E. Vicari and A. Zaffaroni are gratefully
acknowledged. We wish to thank H. Panagopoulos for pointing out
a mistake in the appendix. Special thanks go to I. Klebanov,
 M. L\"uscher,  A. Ritz and M. Strassler for insightful remarks.

The work of M.S. is supported in part  by DOE grant DE-FG02-94ER408.

\section*{Appendix}
\renewcommand{\theequation}{A.\arabic{equation}}
\setcounter{equation}{0}

In this appendix we calculate the distance where two 1-strings merge
into a 2-string (see Fig.~\ref{st1}b). Let us denote the angle between the two
1-strings by $2\alpha$. Then the forces on the junction yield
\beq
\sigma _2 = 2\sigma _1 \cos \alpha\,.
\eeq
Assuming the sine formula \eqref{dsf} for the string tension, we obtain
$$\alpha = {\pi \over N}\,,$$ which is physically reasonable: 
in the large $N$ limit the two strings
hardly interact, and the ``merging angle'' should vanish.

Comparing the energy of the two separate 1-strings (Fig.~\ref{st2}) with the
energy of the configuration in Fig.~\ref{st1}b we arrive at 
\beq
2\sigma _1 L = 4\sigma _1\,  {l\over 2 \sin \alpha} + \sigma _2 \left( L-{l\over
2 \tan \alpha}\right)\,. 
\label{energetics}
\eeq
Now, by substituting the sine formula \eqref{dsf} and the angle $\alpha =
{\pi \over N}$ we obtain
\beq
L\left( 1-\cos {\pi \over N}\right) = l \left ({{2-2\cos {\pi \over N}} \over {2\sin {\pi
\over N}}} \right ).
\eeq
Thus, in the large $N$ limit, the critical distance is
\beq
l = L \, {\pi \over 2N}.
\eeq

\end{document}